# Study of Asymmetric Contact Based Si Wire Solar Cell

M. Golam Rabbani[1], Amit Verma[2], Reza Nekovei[2], Mahmoud M. Khader[3] and M. P. Anantram[1]

[1]Dept. of Electrical Engineering, University of Washington, Seattle, WA, 98195. [2]Dept. of Electrical Engineering and Computer Science, Texas A & M University – Kingsville, Kingsville, TX 78363. [3]Gas Processing Center, College of Engineering, Qatar University, Doha, P.O. 2713, Qatar.

*Abstract* — We simulate single silicon nanowire (SiNW) solar cells with dissimilar work function metal contacts. Both short circuit current ($I_{SC}$) and open circuit voltage ($V_{OC}$) have been investigated. Effects of nanowire dimension, minority carrier lifetime, and contact metal work function difference are understood through simulations. Both $I_{SC}$ and $V_{OC}$ increase with nanowire length but saturate due to minority carrier recombination. The saturation length is found to be five times the diffusion length. The larger the contact work function difference, the more improved the solar cell characteristics. Large work function differences may also avoid need for any doping in axial p-i-n nanowire solar cells. Saturation in $I_{SC}$ as well as degradation in current density with length can be minimized by spreading the contacts along the length of the nanowire.

## I. INTRODUCTION

One dimensional nanowires are promising for photovoltaic applications [1]–[3] and there are many studies on solar cells based on multiple as well as single nanowires already reported in the literature [4]–[8]. Tsakalakos *et al.* [4] and Sivakov *et al.* [5] studied solar cells based on arrays of nanowires and they both achieved moderate to large current densities as well as low optical reflectance. The p-n junction based solar cells in reference [4] were fabricated on metal foils with p-type silicon nanowire arrays covered by n-type amorphous silicon and produced a promising current density of ~1.6mA/cm$^{-2}$ In reference [5], vertical silicon nanowire array based solar cells were fabricated on glass substrates by electroless wet chemical etching, and maximum $V_{OC}$ of 450 mV, short circuit current density ($J_{SC}$) of 40 mA/cm$^{-2}$ and a high power conversion efficiency of 4.4% were obtained. Experimental study [7] on solar cells comprising multiple SiNWs with dual work function metals obtained a high $I_{SC}$ of 91 nA. The relatively fewer studies on single nanowire based solar cells also look promising. Tian *et al.* [6] studied single p-i-n coaxial silicon nanowires and measured open circuit voltage ($V_{OC}$) of 0.26 V and short circuit current ($I_{SC}$) of 0.503 nA while Kelzenberg *et al.* [8] studied single-nanowire solar cells with one rectifying junction created by electrical heating of the segment of the nanowire beneath it. For a nanowire of diameter 900 nm, they achieved a $V_{OC}$ of 0.19 V and a $J_{SC}$ of 5.0 mA/cm$^2$. These experimental works represent first steps towards nanotechnology based third generation solar cells, and further studies are needed to understand the underlying physics, their various limitations as well as to find ways to improving device designs to optimizing photovoltaic performances. In this work, we do device simulation to model a single nanowire. Our goals are to investigate effects of nanowire dimensions, its minority carrier lifetime or diffusion length, difference of metal contact work functions on short circuit photo current as well as open circuit voltage. We also propose a way to improving $I_{SC}$, especially for long wires.

Section II describes the device structure and the simulation approach. Section III discusses the major results while the paper is concluded in Section IV.

## II. DEVICE STRUCTURE AND SIMULATION METHOD

Fig. 1 is a sketch of the device structure under study. Here a rectangular cross-sectional area nanowire is supported on an insulating substrate. Length, width and height of the wire are denoted by L, W and H, respectively. At each of the two ends of the wire, there is a metal pad with a specific work function. Light is incident vertically on the top surface (of area L x W) of the wire. In all simulations, a one nm thick oxide layer is assumed to cover the Si nanowire surface except where there are metal pads. The substrate plays no role in simulation.

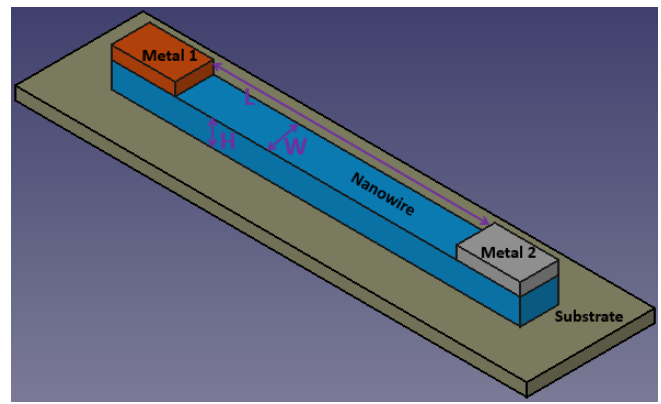

Fig. 1. Schematic representation of a single nanowire device with two dissimilar metal pads. Wire dimensions - length (L), width (W) and height (H) - are indicated, and the downward pointing array of green arrows represents the incident light beam.

Devices simulations were done with Silvaco Atlas [10] software, in which drift-diffusion and carrier continuity equations were solved self-consistently with Poisson's equation. Metal contacted surfaces of the wire are specified with an associated work function whereas in surfaces without any metal contact, homogeneous Neumann boundary conditions are applied. Optical generation is calculated in ray tracing method. To simulate carrier recombination, we have considered both Shockley-Read-Hall and Auger processes.

## III. RESULTS AND DISCUSSION

In this section, we discuss the important results in our work. For all results, the light source is considered to be the standard solar spectrum air mass 1.5 (AM1.5). A doping density of $\sim 10^{15} \text{cm}^{-3}$ (p-type) has been used throughout, and metal work functions of 5.5eV (left contact) and 4.0eV (right contact) have been employed.

### A. Short Circuit Current and Open Circuit Voltage

In Fig. 2, we plot the I-V profile (with light) of a nanowire with W=100 nm and H=35 nm. The length of the wire varies from 10 μm to 1000 μm, and minority carrier lifetime of 10 μsec, which corresponds to a diffusion length of about 160 μm, is used. The photo current versus voltage curves are typical of any solar cells, but note that they move down in current axis as the length of the wire increases. $I_{SC}$ (current at zero voltage) first increases fast with length but eventually saturates at larger lengths. This is because as the length increases, the probability of recombination of excess charge carriers before they reach their respective contacts also increases. $V_{OC}$, as shown in the inset of Fig. 2, has a similar trend. Simulations of wider wires (not shown here) reveals that current scales linearly with wire width.

Fig. 2. Photo circuit current vs applied voltage for a nanowire of 100 nm width and 35 nm thickness. The wire length varies from 10 μm to 1000 μm. The inset shows $V_{OC}$ vs wire length.

### B. Photocurrent vs Lifetime

In Fig. 2, we saw that the photo current saturates for long nanowires. However, as the length of the nanowire increases, the area of its top surface - the illuminated area - increases linearly with length for a fixed width. The larger illuminated area, in turn, means more light absorption and increased photogenerated carrier densities inside the wire for fixed height. Hence we should expect larger photo currents for longer wires. This explains the initial photocurrent increase with length. Saturation, however, occurs due to recombination of minority carriers as they travel towards a metal contact. This is the behavior seen in Fig. 3, which plots the short circuit photo current versus wire lengths for minority carrier lifetimes of 1μsec (solid curve with squares) and 0.1 μsec (dashed curve with circles). The saturation lengths, extracted from Fig. 2, for the two cases are approximately 250 μm and 80 μm, respectively, while the diffusion lengths are 50.8 μm and 16.1 μm, respectively. Thus the saturation lengths are about five times the diffusion length. This can be explained as follows: For a photo generated carrier concentration of $dn_0$ at a location $x_0$, the excess carrier density at location x due to diffusion is

$$dn(x) = dn_0 \exp(-(x-x_0)/L_n) \qquad (1)$$

where $L_n$ is diffusion length. Assuming, a distance $x-x_0=5L_n$,

$$dn(x) = dn_0 \exp(-5) \approx 0 \qquad (2)$$

Thus carriers generated beyond five times the diffusion length from the contact are totally recombined inside the wire, and this causes photo current saturation.

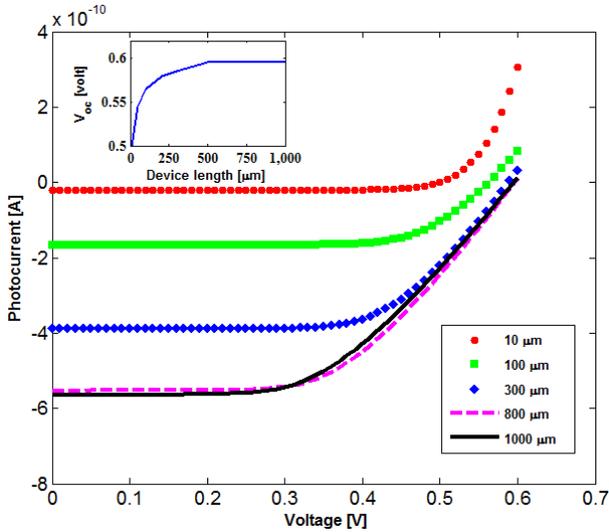

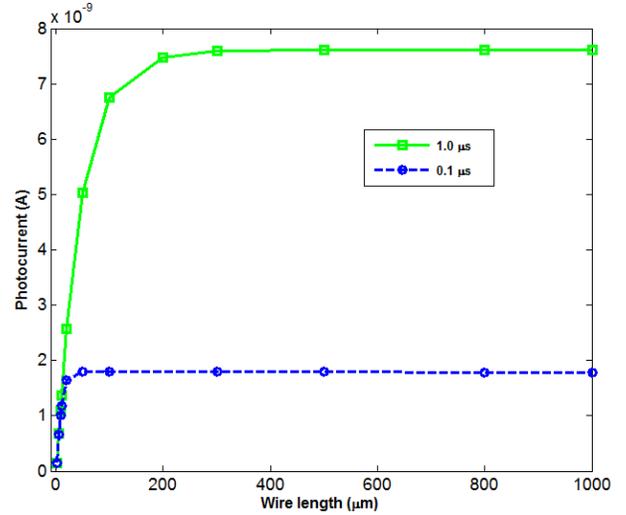

Fig. 3. Photocurrent *vs* wire length for two different minority carrier lifetimes. Lifetimes of 1μsec (solid curve with squares) and 0.1μsec (dashed curve with circles) have been used.

## C. Metal Work Function Difference and Alternative to Doping

From simulations, we find that the larger the difference between the two metal work functions, the larger the contact asymmetry, and the better the photovoltaic properties. This is obvious from the green curves in Fig. 4. Minority carrier lifetime of 10 μsec has been assumed. Here, solid line with circles is for contact work function difference of 0.5eV while solid line with squares is for 1.5eV. Fig. 4 also plots the photocurrent for a p-i-n nanowire device in which the contacts are Ohmic but the parts of the nanowire below the contacts are doped with equal but opposite doping types. p-i-n nanowire solar cells have been used in [11], [12]. However, it is usually difficult to precisely control the doping profile in a nanodevice [13]–[15] so that if the metal work function difference can eliminate the need of any doping, it will be advantageous. Indeed, we see in Fig. 4 that metals with work function difference of 1.5 eV (work functions of 4.0eV and 5.5eV) can give the same short circuit photocurrent as that produced by doping concentrations equal to or in excess of 1E20 $cm^{-3}$. Thus appropriate selection of metals can replace doping requirements and make the design simple and low cost.

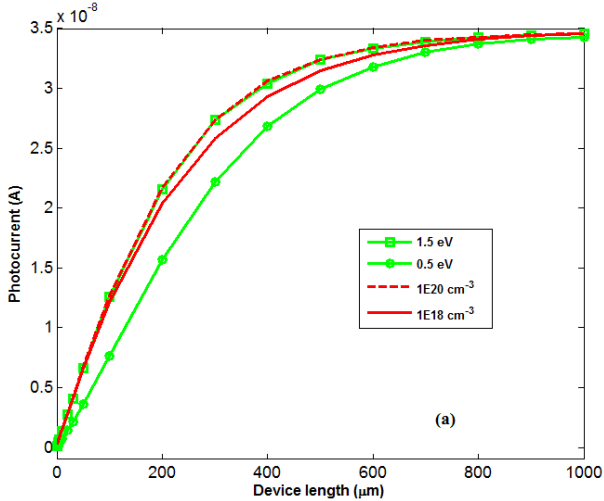

Fig. 4. Short circuit photo current vs wire length for two different work function differences (solid with circles: 0.5 eV and solid with squares: 1.5 eV) and two doping values (solid: 1E18 $cm^{-3}$ and dashed: 1E20 $cm^{-3}$).

## D. Short Circuit Current per Top Surface Area

In addition to $I_{SC}$ and $V_{OC}$, current density is an important parameter for a solar cell as it represents the area efficiency of the device. In a conventional solar cell, current flows from the top of the device to its bottom so that short circuit current divided by the device top surface area gives the short circuit current density. However, in our horizontally lying nanowires current flows parallel to the top surface. So for our device, we define $J_{SC} = I_{SC} / (L*W)$. Fig. 5 plots the $J_{SC}$ for minority carrier lifetimes of 1μsec (solid curve with squares) and 0.1 μsec (dashed curve with circles). $J_{SC}$ decreases as wire length increases, with the shorter the lifetime the faster the decrease. The reason is again attributed to increased recombination of photo generated carriers before collection at the contacts as wire length increases. As already pointed out short circuit current scales linearly with wire width, so the curves in Fig. 5 are independent of wire width.

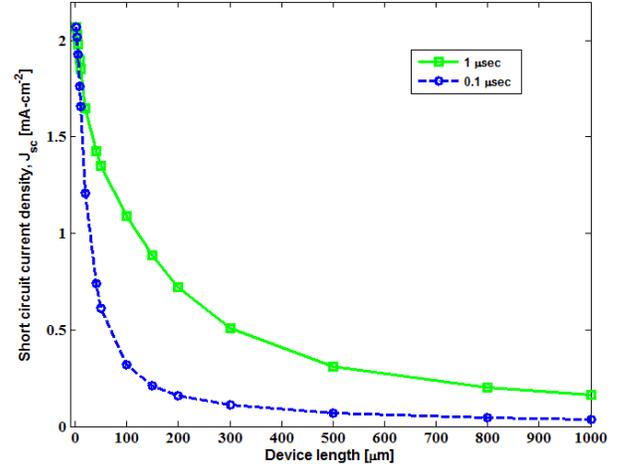

Fig. 5. Short circuit current per nanowire top surface area ($J_{SC}$) vs wire length for three representative minority carrier lifetimes.

## E. Improving Short Circuit Current in Long Wires

For long wires, as seen in Fig. 5, saturation of short circuit current is a drawback since the current per surface area decreases with length. We have researched ways to minimizing the drawback and found that it is possible to make an improvement with some simple design modifications. Instead of having just two metal contacts at the two ends of the wire, we can distribute the contacts along the length of a long wire, with every other contacts shorted together and having only a relatively shorter distance between the consecutive (but opposite) contact pads. This causes the photo generated carrier collection efficiency to increase since electrons and holes now have to travel shorter distances to reach a contact. We found that distance between two consecutive metal pads should be on the order of a diffusion length to gain any improvement in current. The longer the wire the higher the improvement. For a nanowire of length 1000 μm, the short circuit current improves by a factor of more than 3. If the nanowire has lots of surface roughness so that lifetime is smaller, the improvement with multiple contacts will be possible at even shorter lengths as diffusion length will be shorter. This result is very promising and we plan to do further studies to understand it.

## IV. CONCLUSION

A single nanowire with two dissimilar metal contacts have been studied with device simulation. Both short circuit current and open circuit voltage increase with wire length but they eventually saturates due to recombination of minority carriers as they transport towards the contact pads. The saturation length is found to be five times the diffusion length which is understood to be the result of exponential nature of the density of diffused carriers. Effects of metal work function difference on photovoltaic properties have also been studied. The larger the difference, the better the solar cell. Moreover, sork function difference of 1.5eV can eliminate any doping necessary in a p-i-n type axial solar cells. The current density decreases fast with wire length due to the saturation effect. We believe our multiple contact design can minimize such drawbacks.

ACKNOWLEDGEMENT

The work of M. Golam Rabbani and M. P. Anantram was supported by the National Science Foundation under Grant No. 1001174 and by the QNRF grant (NPRP 5 – 968 – 2 – 403). M. P. Anantram was also supported by the University of Washington. Amit Verma, Mahmoud M. Khader, and Reza Nekovei were supported by the QNRF grant (NPRP 5 – 968 – 2 – 403).

REFERENCES

[1] E. Garnett and P. Yang, "Light trapping in silicon nanowire solar cells," *Nano Lett.*, vol. 10, no. 3, pp. 1082–1087, Mar. 2010.

[2] M. M. Adachi, M. P. Anantram, and K. S. Karim, "Core-shell silicon nanowire solar cells," *Sci. Rep.*, vol. 3, p. 1546, Jan. 2013.

[3] Y. Cui, J. Wang, S. R. Plissard, A. Cavalli, T. T. T. Vu, R. P. J. van Veldhoven, L. Gao, M. Trainor, M. a Verheijen, J. E. M. Haverkort, and E. P. a M. Bakkers, "Efficiency enhancement of InP nanowire solar cells by surface cleaning," *Nano Lett.*, vol. 13, no. 9, pp. 4113–4117, Sep. 2013.

[4] L. Tsakalakos, J. Balch, J. Fronheiser, B. a. Korevaar, O. Sulima, and J. Rand, "Silicon nanowire solar cells," *Appl. Phys. Lett.*, vol. 91, no. 23, p. 233117, 2007.

[5] V. Sivakov, G. Andrä, A. Gawlik, A. Berger, J. Plentz, F. Falk, and S. H. Christiansen, "Silicon Nanowire-Based Solar Cells on Glass : Synthesis , Optical Properties , and Cell Parameters," *Nano Lett.*, vol. 9, no. 4, pp. 1549–1554, 2009.

[6] B. Tian, X. Zheng, T. J. Kempa, Y. Fang, N. Yu, G. Yu, J. Huang, and C. M. Lieber, "Coaxial silicon nanowires as solar cells and nanoelectronic power sources," *Nature*, vol. 449, no. 7164, pp. 885–890, Oct. 2007.

[7] J. Kim, J.-H. Yun, C.-S. Han, Y. J. Cho, J. Park, and Y. C. Park, "Multiple silicon nanowires-embedded Schottky solar cell," *Appl. Phys. Lett.*, vol. 95, no. 14, p. 143112, 2009.

[8] M. D. Kelzenberg, D. B. Turner-Evans, B. M. Kayes, M. a Filler, M. C. Putnam, N. S. Lewis, and H. a Atwater, "Photovoltaic measurements in single-nanowire silicon solar cells," *Nano Lett.*, vol. 8, no. 2, pp. 710–714, Feb. 2008.

[9] F. Zhang, T. Song, and B. Sun, "Conjugated polymer-silicon nanowire array hybrid Schottky diode for solar cell application," *Nanotechnology*, vol. 23, no. 19, p. 194006, May 2012.

[10] "Atlas, Silvaco. Available: http://www.silvaco.com," 2014. [Online]. Available: http://www.silvaco.com/.

[11] T. J. Kempa, B. Tian, D. R. Kim, J. Hu, X. Zheng, and C. M. Lieber, "Single and tandem axial p-i-n nanowire photovoltaic devices.," *Nano Lett.*, vol. 8, no. 10, pp. 3456–60, Oct. 2008.

[12] H. P. T. Nguyen, Y.-L. Chang, I. Shih, and Z. Mi, "InN p-i-n Nanowire Solar Cells on Si," *IEEE J. Sel. Top. Quantum Electron.*, vol. 17, no. 4, pp. 1062–1069, Jul. 2011.

[13] P. Peercy, "The drive to miniaturization," *Nature*, vol. 406, no. 6799, pp. 1023–1026, Aug. 2000.

[14] C. Claeys, "Technological Challenges of Advanced CMOS Processing and Their Impact on Design Aspects," *Proc. 17th Int. Conf. VLSI Des.*, p. 275, 2004.

[15] S. Xiong and J. Bokor, "Structural Optimization of SUTBDG Devices for Low-Power Applications," *IEEE Trans. Electron Devices*, vol. 52, no. 3, pp. 360–366, Mar. 2005.